\documentclass[prd,twocolumn,amsmath,amssymb]{revtex4}
\usepackage{graphicx}% Include figure files

\begin{document}

\title{\boldmath
 Direct measurement of the absolute branching fraction for 
$D^+ \rightarrow {\overline K}^0\mu^+\nu_\mu$ and determination of
$ \Gamma (D^0 \rightarrow K^-\mu^+\nu_\mu)
/ \Gamma (D^+ \rightarrow {\overline K}^0\mu^+\nu_\mu)$ }
%\date{ }
\author{
M.~Ablikim$^{1}$,              J.~Z.~Bai$^{1}$,               Y.~Ban$^{12}$,
X.~Cai$^{1}$,                  H.~F.~Chen$^{16}$,
H.~S.~Chen$^{1}$,              H.~X.~Chen$^{1}$,              J.~C.~Chen$^{1}$,
Jin~Chen$^{1}$,                Y.~B.~Chen$^{1}$
% ...
\\
\vspace{0.2cm}
(BES Collaboration)\\
\vspace{0.2cm}
{\it
$^{1}$ Institute of High Energy Physics, Beijing 100049, People's Republic of China\\
% ...
$^{12}$ Peking University, Beijing 100871, People's Republic of China\\
% ...
$^{16}$ University of Science and Technology of China, Hefei 230026, People's Republic of China\\
% ...
}}

\begin{abstract}
   The absolute branching fraction for the decay 
$D^+ \rightarrow {\overline K}^0\mu^+\nu_\mu$ is determined using $5321 \pm 149 \pm160$
singly tagged $D^-$ sample from the data collected around 3.773 GeV with
the BESII detector at the BEPC collider. In the system recoiling against
the singly tagged $D^-$ mesons, $28.7 \pm 6.4$ events for
$D^+ \rightarrow {\overline K}^0\mu^+\nu_\mu$ are observed. These yield
the absolute branching fraction to
be $BF(D^+ \rightarrow {\overline K}^0\mu^+\nu_\mu) = (10.3\pm2.3\pm0.8)\%$. 
The ratio of the two partial widths for the decays $D^0 \rightarrow K^-\mu^+\nu_\mu$ 
and $D^+ \rightarrow {\overline K}^0\mu^+\nu_\mu$ is determined to be
$\Gamma (D^0 \rightarrow K^-\mu^+\nu_\mu)/ 
\Gamma (D^+ \rightarrow {\overline K}^0\mu^+\nu_\mu) = 0.87 \pm 0.24 \pm 0.15$.
\end{abstract}

\maketitle

\section{\bf Introduction} 
Measurements of the absolute branching fractions for
exclusive semileptonic decays
of charmed mesons can provide important information about decay
mechanism of the mesons. 
If the isospin symmetry holds
in the exclusive semileptonic decays of the charged and neutral $D$ mesons, 
the two partial widths for the decays
$D^0\rightarrow K ^-l^+\nu_l$ and 
$D^+ \rightarrow {\overline K}^0 l^+\nu_l$ are expected to be equal, which means
the ratio 
$\Gamma (D^0 \rightarrow K^-l^+\nu_l)/
\Gamma (D^+ \rightarrow {\overline K}^0 l^+\nu_l)$ = 1.
The measured branching fractions for the semielectronic
decays historically yielded the ratio
$\Gamma (D^0 \rightarrow K^-e^+\nu_e)/
\Gamma (D^+ \rightarrow {\overline K}^0 e^+\nu_e) = 1.4\pm 0.2$\cite{shipsey}, 
which deviates
from 1.0 by $2 \sigma$. This was historically called a "long-standing puzzle"
in the $D$ exclusive semilelectronic decays. Recently, BES
Collaboration measured the ratio 
$\Gamma (D^0 \rightarrow K^-e^+\nu_e)/
\Gamma (D^+ \rightarrow {\overline K}^0 e^+\nu_e) = 1.08\pm
0.22\pm0.07$~\cite{dptok0ev_bes}, 
indicating that the isospin symmetry holds in the exclusive semielectronic
decays of the charged and neutral $D$ mesons,
and thereby solving the "long-standing puzzle"~\cite{shipsey}. 
This was confirmed by the 
CLEO measurement~\cite{cleo_dptok0ev}.

  In this Letter, we report a direct measurement of the absolute branching
fraction for the decay $D^+ \rightarrow {\overline K}^0\mu^+\nu_\mu $
and determination of the ratio of the two partial widths
for the semimuonic decays
${\Gamma(D^0 \rightarrow {K}^-\mu^+\nu_\mu})/
{\Gamma(D^+ \rightarrow {\overline K}^0\mu^+\nu_\mu})$, 
which can be used to check
if the isospin symmetry holds in the exclusive semimuonic decays.

\section{\bf The BESII DETECTOR}

The BESII is a conventional cylindrical magnetic detector that is
described in detail in Ref.~\cite{bes}.  A 12-layer vertex chamber
(VC) surrounding the beryllium beam pipe provides input to the event
trigger, as well as coordinate information.  A forty-layer main drift
chamber (MDC) located just outside the VC yields precise measurements
of charged particle trajectories with a solid angle coverage of $85\%$
of $4\pi$; it also provides ionization energy loss ($dE/dx$)
measurements which are used for particle identification.  Momentum
resolution of $1.7\%\sqrt{1+p^2}$ ($p$ in GeV/c) and $dE/dx$
resolution of $8.5\%$ for Bhabha scattering electrons are obtained for
the data taken at $\sqrt{s}=3.773$ GeV. An array of 48 scintillation
counters surrounding the MDC measures the time of flight (TOF) of
charged particles with a resolution of about 180 ps for electrons.
Outside the TOF, a 12 radiation length, lead-gas barrel shower counter
(BSC), operating in self-quenching streamer mode,
measures the energies of
electrons and photons over $80\%$ of the total solid angle with an
energy resolution of $\sigma_E/E=0.22/\sqrt{E}$ ($E$ in GeV) and spatial
resolutions of
$\sigma_{\phi}=7.9$ mrad and $\sigma_Z=2.3$ cm for
electrons. A solenoidal magnet outside the BSC provides a 0.4 T
magnetic field in the central tracking region of the detector. Three
double-layer muon counters instrument the magnet flux return, and serve
to identify muons of momentum greater than 500 MeV/c. They cover
$68\%$ of the total solid angle.

\section{ DATA ANALYSIS}

The data used in the
analysis were collected with the BESII detector at the BEPC collider.
A total integrated luminosity of about 33 $\rm pb^{-1}$ was taken at and
around the center-of-mass energy of $\sqrt{s}=3.773$ GeV. 
Around these energies,
the $\psi(3770)$ resonance is produced in $e^+e^-$ annihilation.
It decays to $D\bar D$ pairs ($D^0\bar D^0$ and $D^+D^-$) 
with a large branching fraction of about $(85 \pm 6)\%$~\cite{bf_psipp_to_dd}.
Taking the advantage of the $\psi(3770)$ decay to $D^+D^-$ pairs, 
we can do absolute measurement of the branching fraction for
$D^+ \rightarrow {\overline K}^0\mu^+\nu_{\mu}$ with a singly tagged $D^-$ sample.
If a $D^-$ meson is fully reconstructed (it is called a singly 
tagged $D^-$ meson) from this data sample, the $D^+$ meson must exist in
the system recoiling against the singly
tagged $D^-$ meson. In the recoil system,
we can select the semileptonic decay 
$D^+\rightarrow {\overline K}^0\mu^+ \nu_\mu$ 
based on the kinematic signature of the singly tagged $D^-$ event, 
and measure the branching fraction for this decay directly.

\subsection{ Events selection}

In order to ensure
the well-measured 3-momentum vectors and the reliably charged particle
identification, the charged tracks used in the single tag analysis
are required to be within $|{\rm cos}\theta|<$0.85,
where $\theta$ is the polar angle of the charged track.
All tracks, save those from $K^0_S$ decays, must originate
from the 
interaction region: the closest approach of the charged
track in the $xy$ plane is less than 2.0 cm and the absolute $z$ position of the
track is less than 20.0 cm.
Pions and kaons are identified by means of
TOF and $dE/dx$ measurements. Pion identification requires a consistency
with the pion hypothesis at a confidence level ($CL_{\pi}$) greater than
$0.1\%$.
In order to reduce misidentification, a kaon candidate is
required to have a larger confidence level ($CL_{K}$) for a kaon hypothesis
than that for a pion hypothesis.
The $\pi^0$ is reconstructed in the decay of
$\pi^0 \rightarrow \gamma\gamma$.
To select good photons from the decay
of $\pi^0$, the energy of a photon deposited in the BSC
is required to be greater than $0.07$ GeV~\cite{d0tokev_bes,dptok0ev_bes,dptomuv_bes}, 
and the electromagnetic shower
is required to start in the first 5 
readout layers of the BSC. 
In order to reduce
backgrounds due to fake photons, 
the angle between the
photon and the nearest charged track is required to be greater
than $22^{\circ}$~\cite{d0tokev_bes}
and the angle between the direction of the cluster development
and the direction of the photon emission to be less than
$37^{\circ}$~\cite{d0tokev_bes}.

For the single tag modes of $D^- \rightarrow K^+\pi^+\pi^-\pi^-\pi^-$ and
$D^- \rightarrow \pi^+\pi^-\pi^-$, backgrounds are further reduced by
requiring the difference between the measured energy
of the $D^-$ candidate and the beam energy
to be less than 70 and 60 MeV, respectively.
In addition, the cosine of the $D^-$ production angle relative to the beam
direction is required to be $|{\rm cos}\theta_{D^-}|<0.8$.

\subsection{Singly tagged $D^-$ sample}

The $D^-$ meson is reconstructed in the
nine hadronic decay modes of
$K^+\pi^-\pi^-$, $K^0\pi^-$, $K^0K^-$, $K^+K^-\pi^-$,
$K^0\pi^-\pi^-\pi^+$, $K^0\pi^-\pi^0$,  $K^+\pi^-\pi^-\pi^0$,
$K^+\pi^+\pi^-\pi^-\pi^-$ and $\pi^+\pi^-\pi^-$.
Events which contain at least
three reconstructed charged tracks with good helix fits are selected.
In order to improve the momentum resolution and 
reduce the combinatorial background in the invariant mass spectra, 
each combination is subject to a one-constraint (1C) kinematic fit requiring
overall event energy conservation and that the unmeasured recoil system has
the same invariant mass as the track combinations.
Candidates with a fit probability $P(\chi^2)$ greater
than $0.1\%$ are retained.
If more than one combination
satisfies
$P(\chi^2)>0.1\%$,
the combination with the largest fit probability is retained.
For the single tag modes
with a neutral kaon and/or neutral pion,
one additional constraint kinematic fit
for the $K^0_S \rightarrow \pi^+\pi^-$ and/or
$\pi^0 \rightarrow \gamma\gamma$ hypothesis is performed, respectively.

\begin{figure}[hbt]
\includegraphics*[width=8.8cm,height=9.0cm]
{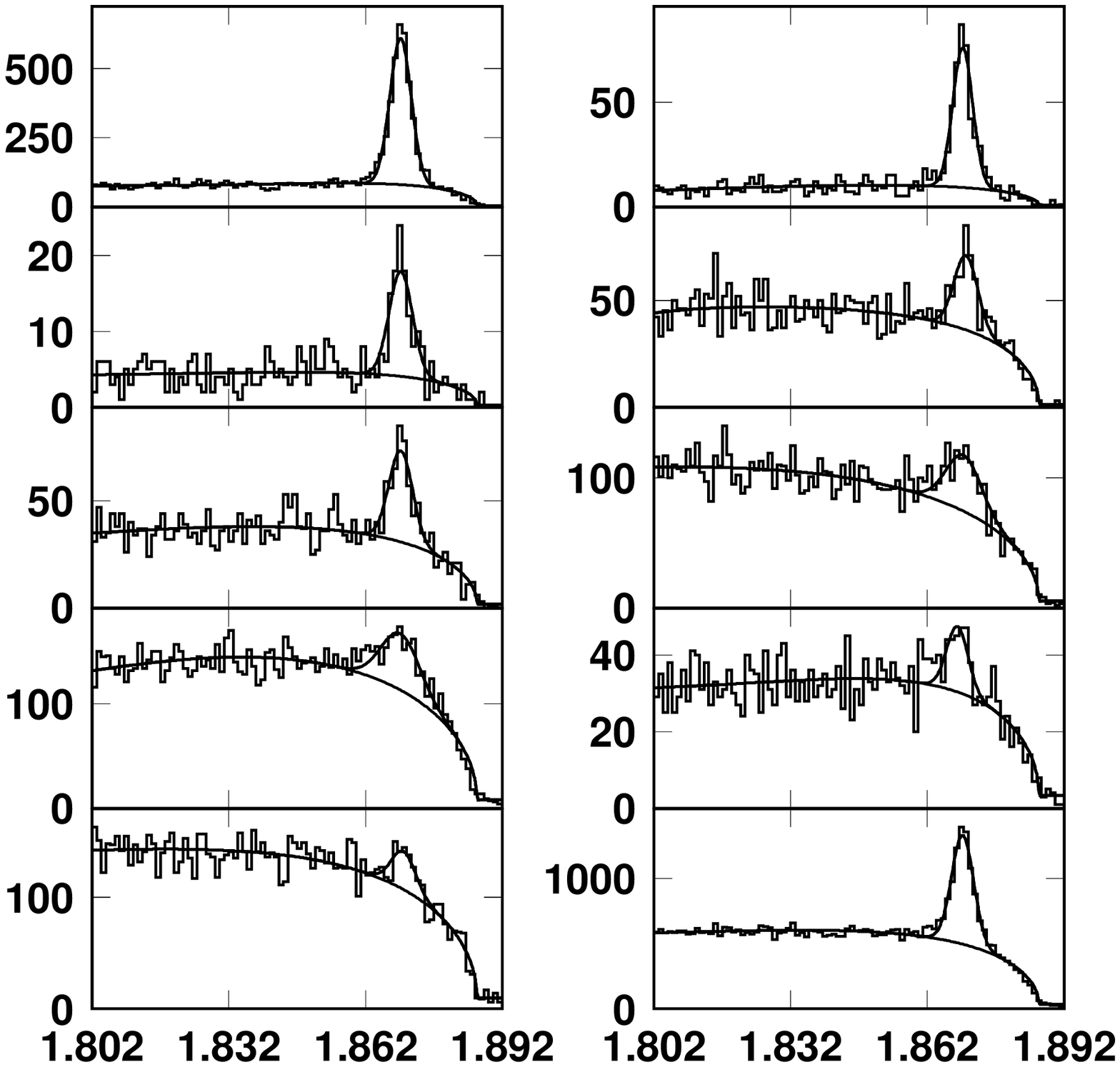}
\put(-195,-6){\bf\large Fitted Mass~~$M_{mKn\pi}$ (GeV/$c^2$)}
\put(-250,60){\rotatebox{90}{\bf\large Events/(0.001 GeV/$c^2$)}}
\put(-200,230){(a)}
\put(-90,230){(b)}
\put(-200,190){(c)}
\put(-90,190){(d)}
\put(-200,125){(e)}
\put(-90,125){(f)}
\put(-200,80){(g)}
\put(-90,80){(h)}
\put(-200,40){(i)}
\put(-90,40){(j)}
\caption{The distributions of the fitted masses of 
(a) $K^+ \pi ^-\pi^-$, (b) $K^0 \pi ^-$, 
(c) $K^0 K ^-$, (d) $K^+K^-\pi^-$, (e) $K^0\pi^-\pi^-\pi^+$,
(f) $K^0\pi ^-\pi ^0 $, (g) $K^+\pi ^-\pi^-\pi ^0$, 
(h) $K^+\pi^+\pi^-\pi^-\pi^-$, (i)$\pi^+\pi ^-\pi^-$ combinations;
(j) is the superposition of the fitted masses of the $mKn\pi$ 
combinations for the nine modes; the
histograms are the data and the solid lines are the fits to the data.
}
\label{dptag}
\end{figure}

Figure \ref{dptag} shows the resulting distributions of the fitted
invariant masses of the $mKn\pi$ ($m=0,1,2,n=0,1,2,3,4$)  combinations,
which are calculated using the fitted momentum vectors from the kinematic
fit. 
The signals for the singly tagged $D^-$ mesons are clearly observed
in the fitted mass spectra.
A maximum likelihood fit to the mass
spectrum with a Gaussian function for the $D^-$ signal and a special
function~\cite{dptok0ev_bes}~\cite{dptomuv_bes} to describe the background shape, yields
the observed numbers of the singly tagged $D^-$ mesons for each 
of the nine modes and the total number of $5321\pm 149 \pm 160$
reconstructed $D^-$ mesons, where the
first error is statistical and the second systematic obtained
by varying the parameterization of the background.

\subsection{Candidates for $D^+ \to {\overline K}^0\mu^+\nu_\mu$}
The candidates for the decay $D^+ \to {\overline K}^0\mu^+\nu_\mu$ are 
selected from the tracks in the recoil system.
The candidates are only allowed to have three good
charged tracks in the recoil side. 
One of them is identified as muon with the charge opposite to
that of the tagged $D^-$ 
and the other two are identified as $\pi^+$ and $\pi^-$.
The $\pi^+$ and $\pi^-$ are identified by requiring the confidence level
$CL_{\pi}$ for the pion hypothesis to be greater than the confidence level 
$CL_{K}$ for the kaon hypothesis or by requiring $CL_{\pi}>0.1\%$. 
To select the ${K}^0_S$, it is required that
the invariat mass of the $\pi^+ \pi^-$ combination 
should be within $\pm 20$ MeV/$c^2$ mass window 
of the ${K}^0_S$ nominal mass and
the $\pi^+\pi^-$ must originate from a secondary    
vertex which is displaced 
from the event vertex at least by 4 mm.

To obtain the information about the missing neutrino from the
semileptonic decay $D^+ \to {\overline K}^0\mu^+\nu_\mu$ under study,
a kinematic quantity $U_{miss} = E_{miss} - p_{miss}$
is defined, where $E_{miss}$ and $p_{miss}$ are the total energy and
momentum of all missing particles in the event.
To select candidates for $D^+ \rightarrow {\overline K}^0\mu^+\nu_\mu$,
it is required that each event should have its 
$|U_{miss,i}| <2\sigma_{U_{miss,i}}$, where $\sigma_{U_{miss,i}}$ is the standard
deviation of the $U_{miss,i}$ distribution, which is obtained by analyzing the
Monte Carlo events of $D^+ \to {\overline K}^0\mu^+\nu_\mu$
versus the $i$th singly tagged $D^-$ mode.
 
\begin{figure}
\begin{center}
 \includegraphics*[width=9.0cm,height=9.0cm] 
{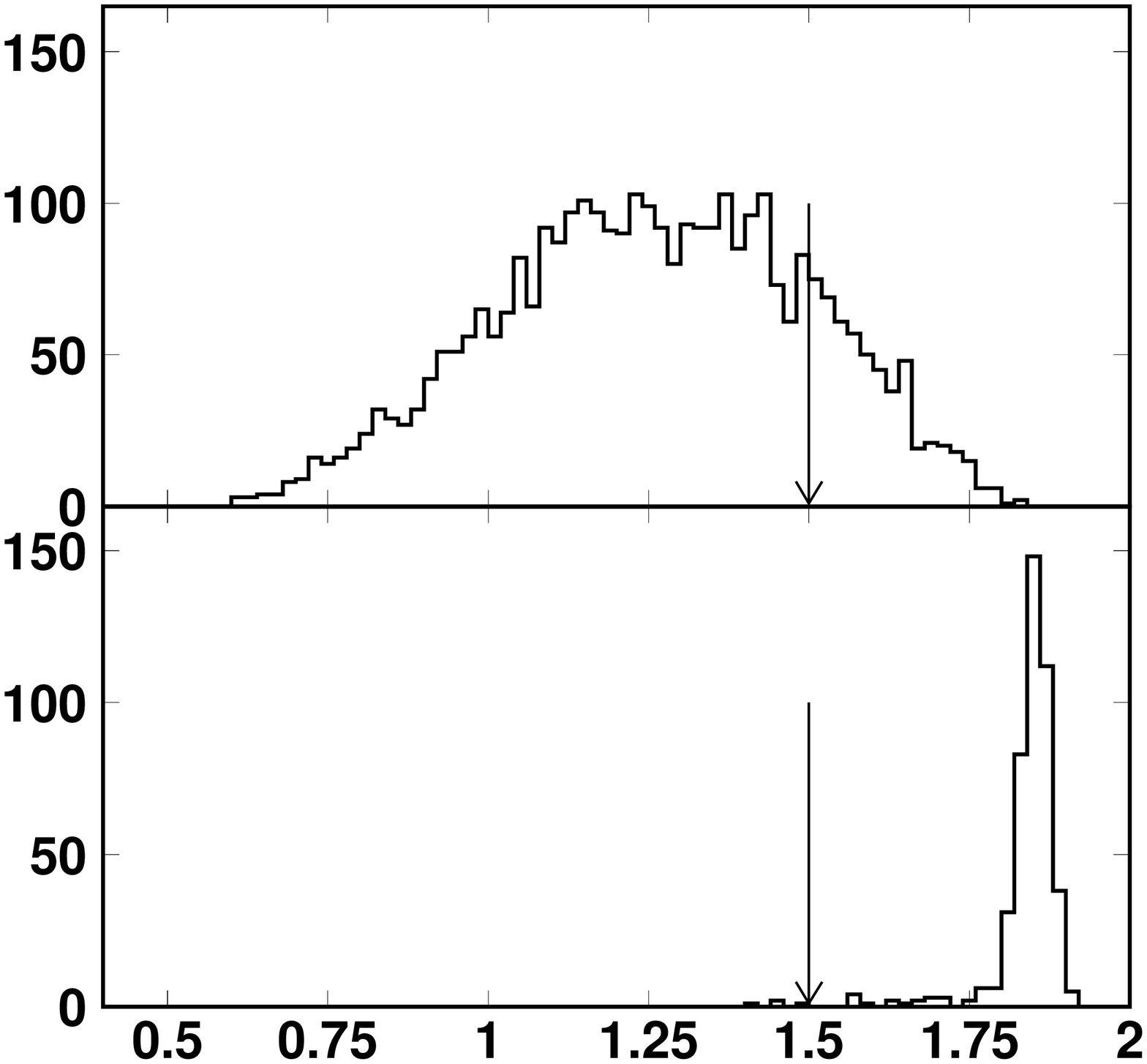}
\put(-205,-3){\bf\large Fitted Mass~~$M_{{\overline K^0}\mu^+}$ (GeV/$c^2$)}
\put(-260,75){\rotatebox{90}{\bf\large Events/(0.02 GeV/$c^2$)}}
\put(-200,210){(a)}
\put(-200,120){(b)}
\caption{The distributions of the invariant masses of the ${\overline K}^0\mu^+$ 
combinations for the Monte Carlo events of (a)
$D^+ \rightarrow {\overline K}^0\mu^+\nu _{\mu} $ versus $D^- \to K^+\pi^-\pi^-$;
(b) $D^+ \rightarrow {\overline K}^0\pi^+$ versus $D^- \to K^+\pi^-\pi^-$,
where the ${\overline K}^0\pi^+$ are misidentified as ${\overline K}^0\mu^+$. 
}
\label{rec_cut_k0bmu}
\end{center}
\end{figure}

\begin{figure}
\includegraphics*[width=9.0cm,height=5.0cm]
{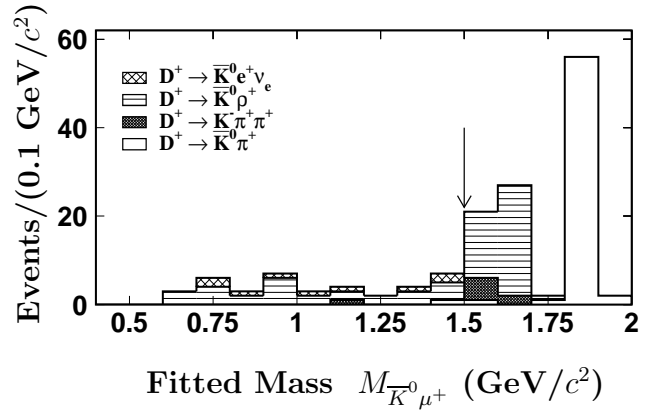}
\put(-205,-6){\bf \large Fitted Mass~~$M_{{\overline K}^0\mu^+}$ (GeV/$c^2$)}
\put(-255,20){\rotatebox{90}{\bf\large Events/(0.1 GeV/$c^2$)}}
\caption{The distribution of the invariant masses of ${\overline K}^0\mu^+$
combinations for the Monte Carlo events of
$e^+e^- \rightarrow D\bar D$ ($D^0\bar D^0$ and $D^+D^-$), where the
$D$ mesons are set to decay into all possible final states except for
$D^+ \rightarrow {\overline K}^0\mu^+\nu _{\mu}$ (see text).
}
\label{rec_cut}
\end{figure}

The main sources of the backgrounds to the semileptonic decay are 
$D^+ \rightarrow {\overline K}^0\pi^0\mu^+\nu_\mu$
and $D^+ \rightarrow  {\overline K}^0\pi^+\pi^0$. These backgrounds
can be suppressed by rejecting the events with extra isolated photons which
are not used in the reconstruction of the singly tagged $D^-$.
The isolated photon should have its energy 
to be greater than 0.1 GeV and should satisfy photon selection criteria 
as mentioned earlier.
In the data analysis, we do not use the information from the muon counter to
separate the muon from pion since we want to reconstruct more signal events
for the semileptonic decay. Due to misidentifying a pion (or an electron) as a muon, 
the events such as $D^+ \rightarrow {\overline K}^0\pi^+$, 
$D^+ \rightarrow K^- \pi^+\pi^+$,
$D^+ \rightarrow {\overline K}^0 \rho^+$ and
$D^+ \rightarrow {\overline K}^0 e^+\nu_e$
could be misidentified as $D^+ \to {\overline K}^0\mu^+\nu_\mu$.
Monte Carlo study shows that these events are the
main backgrounds to the semileptonic decay $D^+ \to {\overline K}^0\mu^+\nu_\mu$.
However, these background events can be suppressed by requiring
the invariant masses of $\overline K^0 \mu^+$ combinations 
to be less than 1.5 GeV/$c^2$.
Figure~\ref{rec_cut_k0bmu} shows the distribution of the invariant masses of
${\overline K}^0 \mu^+$ combinations from the Monte Carlo events of
$e^+e^- \rightarrow \psi(3770) \rightarrow D^+D^-$, where
figure~\ref{rec_cut_k0bmu}(a) shows the distribution for the events of
$D^-\rightarrow K^+\pi^-\pi^-$ versus 
$D^+ \rightarrow {\overline K}^0\mu^+\nu_{\mu}$ 
and figure~\ref{rec_cut_k0bmu}(b)
shows the distribution
for the events of
$D^-\rightarrow K^+\pi^-\pi^-$ versus $D^+ \rightarrow {\overline K}^0\pi^+$. 
The invariant masses of ${\overline K}^0 \mu^+$ combinations are widely
distributed from 0.6 to 1.8 GeV/$c^2$ due to missing of the neutrino from the
semileptonic decay $D^+ \rightarrow {\overline K}^0\mu^+\nu_{\mu}$, 
while the invariant masses of ${\overline K}^0 \mu^+$ combinations misidentified
from the decay $D^+ \rightarrow {\overline K}^0\pi^+$ 
are concentrated at the peak of the $D^+$ mass.
Figure~\ref{rec_cut} shows the distribution
of the invariant masses of ${\overline K}^0\mu^+$
combinations for the Monte Carlo events of $e^+e^- \rightarrow D\bar D$
($D^0\bar D^0$ and $D^+D^-$),
where the $D\bar D$ are set to decay into all possible final states according to
their decay modes and branching fractions quoted from PDG06~\cite{pdg06}.
These selected events are 
misidentified from the source of the main backgrounds as mentioned above
and marked on the figure. The Monte Carlo sample is fourteen times larger
than the data sample.
The criterion $M_{{\overline K}^0 \mu^+}<1.5$ GeV/$c^2$ can reject most of the
main backgrounds to the selected semileptonic decay events.

\begin{figure}
\includegraphics*[width=9.0cm,height=5.0cm]
{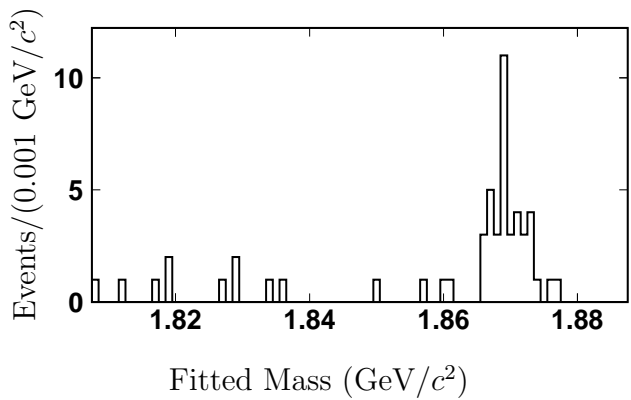}
\put(-195,-6){\large Fitted Mass (GeV/$c^2$)}
\put(-255,20){\rotatebox{90}{\large Events/(0.001 GeV/$c^2$)}}
\caption{The distribution of the fitted masses of $mKn\pi$
combinations for the events for which the 
$D^+ \rightarrow {\overline K}^0 \mu^+\nu_{\mu}$ candidates are observed in the
system recoiling against the $mKn\pi$ combinations.
}
\label{dpmu}
\end{figure}

Figure~\ref{dpmu} shows the distribution of the fitted invariant masses
of the $mKn\pi$ combinations for the events for which the 
$D^+ \rightarrow {\overline K}^0\mu^+\nu_{\mu}$ candidates
satisfing the selection criteria mentioned above are 
observed in the system recoiling against the $mKn\pi$ combinations,
where a clear signal for the singly tagged $D^-$ is observed.
In Fig.~\ref{dpmu}, there are 38 events in the $\pm 3\sigma_{mass,i}$ 
signal regions, while there are 15 events in the outside of the signal
regions; where the $\sigma_{mass,i}$ is the standard deviation of the 
fitted mass distribution for the single tag mode($i$) ($i=1$ is for 
$K^+\pi^-\pi^-$; $i=2$ is for $K^0\pi^-$... and $i=9$ is for
$\pi^+\pi^-\pi^-$ modes). Assuming that
the background distribution is flat except the ones described in subsection
D, 5.2$\pm$1.4 background events are
estimated in the signal region. 
In addition, there may also be the $\pi^+\pi^-$ combinatorial background.
By selecting the events in which the invariant masses of the $\pi^+\pi^-$
combinations in the recoil side of the tags are outside of the $K^0_S$ mass
window, we estimate that there are $0.6 \pm 0.3$ background events in the
candidate events.
After subtracting these numbers of backgrounds
we obtain $32.2 \pm 6.3$ candidates for   
$D^+\rightarrow {\overline K}^0\mu^+\nu_\mu$ decay.

\vspace{1.2cm}
\begin{figure}                              
\begin{center}                                              
\includegraphics[width=9.0cm,height=5.0cm]
{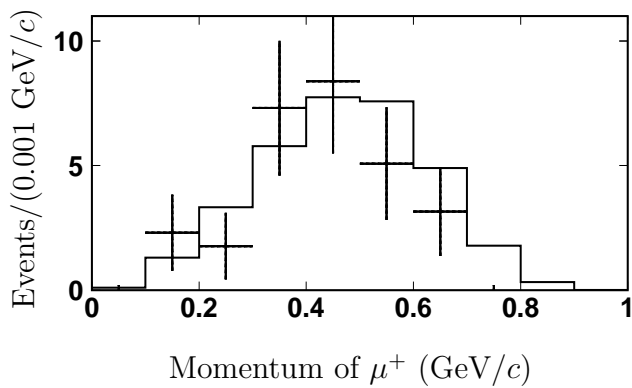}
\put(-195,-6){\large Momentum of $\mu^+$ (GeV/$c$)}
\put(-255,20){\rotatebox{90}{\large Events/(0.001 GeV/$c$)}}
\caption{The distribution of the muon momenta from the selected candidates
for $D^+ \rightarrow {\overline K}^0\mu^+\nu _{\mu}$ decay,
where error bars are the data after subtracting the backgrounds
estimated from Monte Carlo $D\bar D$ events, 
while the histogram is from the Monte Carlo events for 
$D^+ \rightarrow {\overline K}^0\mu^+\nu _{\mu}$.
}                                         
\label{dpmu_momentum}
\end{center}
\end{figure}

\subsection{Background Subtraction}
  
  There are, however, still some background contaminations 
in the selected candidate events due to some other semileptonic
decays or hadronic decays, which are not rejected by the selection criteria
as described above. These background events must be subtracted 
from the sample of the selected candidate events. The numbers of background
events can be estimated by analyzing the Monte Carlo sample of
fourteen times larger than the data. 
By analyzing this Monte Carlo sample and normalize
the number of the background events observed from this sample,
totally $3.5 \pm 0.7$ background events are obtained 
in the selected semileptonic decay sample from the data.
After subtracting the number of background events,
$28.7 \pm 6.4$ signal events for 
$D^+ \rightarrow {\overline K}^0\mu^+\nu_\mu$ decay are retained.
  
Figure~\ref{dpmu_momentum} 
shows the distribution of the
momenta of the $\mu^+$ from the selected candidate events for the decay
$D^+ \rightarrow {\overline K}^0\mu^+\nu_{\mu}$, respectively. The error bars in the
figure show the events from the data after subtracting the backgrounds 
estimated from the Monte Carlo $D\bar D$ events, 
while the histogram shows the events from the
Monte Carlo for the decay $D^+ \rightarrow {\overline K}^0\mu^+\nu_{\mu}$.

\section {\bf Results}

The detection efficiency for reconstruction of the semilpetonic decay
$D^+ \rightarrow {\overline K}^0\mu^+\nu_\mu$
is obtained from Monte Carlo simulation. 
The efficiency is 
$\epsilon_{D^+ \rightarrow {\overline K}^0\mu^+\nu_\mu}=
(5.24 \pm 0.06)\%$ including the branching fraction for the decay
$K^0_S \rightarrow \pi^+\pi^-$, where the error is statistical.

The branching fraction is obtained by dividing the observed number
of the semileptonic decay events N($D^+\rightarrow {\overline K}^0\mu^+\nu_\mu$)
by the number of the singly tagged $D^-$ mesons $N_{D^-_{tag}}$ and the 
reconstruction efficiency $\epsilon_{D^+\rightarrow {\overline K}^0\mu^+\nu_\mu}$,
\begin{equation}
    BF(D^+\rightarrow {\overline K}^0\mu^+\nu_\mu) = \frac{N(D^+\rightarrow
{\overline K}^0\mu^+\nu_\mu)}{N_{D^-_{tag}} \times
\epsilon_{D^+\rightarrow {\overline K}^0\mu^+\nu_\mu}}.
\end{equation}
Inserting these numbers in equation (1), we obtain the branching fraction for
$D^+\rightarrow {\overline K}^0\mu^+\nu_\mu$ decay to be
\begin{equation}
BF(D^+ \rightarrow {\overline K}^0\mu^+\nu_\mu)= (10.3\pm2.3\pm0.8)\%,
\end{equation}
where the first error is statistical and the second systematic.
The systematic uncertainty arises
from the particle identification ($\sim 1.8\%$), tracking
efficiency ($\sim 2.0\%$ per track), photon reconstruction ($\sim 2.0\%$), $U_{miss}$
selection ($\sim 0.6\%$), the number of the singly tagged 
$D^-$ ($\sim 3.0\%$),
background fluctuation ($\sim 2.4\%$), uncertainty in background estimation due
to unknown branching fractions of some background channels ($\sim 2.4\%$), and
Monte Carlo statistics($\sim 1.1\%$). Adding these uncertainties
in quadrature yields the total systematic error of $\sim 8.1\%$. 

With the same data sample, BES Collaboration measured 
the absolute branching fraction for $D^0 \rightarrow K^-\mu^+\nu_\mu$ 
decay to be 
$BF(D^0 \rightarrow K^-\mu^+\nu_\mu)=(3.55\pm0.56\pm0.59)\%$~\cite{d0tomu}. 
Using the measured branching fractions for the decays
$D^0 \rightarrow K^-\mu^+\nu_\mu$, 
$D^+ \rightarrow {\overline K}^0\mu^+\nu_\mu$ 
and the lifetimes of the $D^0$ and $D^+$
quoted from the PDG~\cite{pdg06}, 
we determine the ratio of the decay widths
\begin{equation}
\frac{\Gamma(D^0 \rightarrow K^-\mu^+\nu_\mu)}
{\Gamma(D^+ \rightarrow {\overline K}^0\mu^+\nu_\mu)} = 
0.87\pm0.24\pm0.15,
\end{equation}
where the first error is statistical and the second systematic arising
from some uncancelled systematic uncertainty ($\sim 16.6\%$) in the measured ratio
of the branching fractions for the two decays and the uncertainty
($0.8\%$) in the measured ratio of the $D^0$ and $D^+$ lifetimes.

\section {\bf Summary } 
In summary, from analyzing the data sample of about 33 ${\rm pb^{-1}}$ collected
at and around 3.773 GeV with the BESII detector at the BEPC collider,
we measured the branching fraction for the decay 
$D^+ \rightarrow {\overline K}^0\mu^+\nu_\mu$ to be
$BF(D^+ \rightarrow {\overline K}^0\mu^+\nu_\mu) = (10.3\pm2.3\pm0.8)\%$.
Using the values of the measured branching fractions for the decays
$D^0 \rightarrow K^-\mu^+\nu_\mu$ and 
$D^+ \rightarrow {\overline K}^0\mu^+\nu_\mu$, we determined the ratio of the
two partial widths 
$\Gamma(D^0\rightarrow K^-\mu^+\nu_\mu)/
\Gamma(D^+\rightarrow {\overline K^0}\mu^+\nu_\mu) = 0.87\pm 0.24 \pm 0.15$, 
which is consistent within error with
the spectator model prediction. This result supports that the 
isospin conservation holds in the exclusive semimuonic decays of the 
$D^+ \rightarrow {\overline K}^0\mu^+\nu_\mu$ and 
$D^0 \rightarrow K^-\mu^+\nu_\mu$. 

\section {\bf Acknowledgements }

   The BES collaboration thanks the staff of BEPC for their hard efforts.
This work is supported in part by the National Natural Science Foundation
of China under contracts Nos. 10491300, 10225524, 10225525, the Chinese Academy 
of Sciences under contract No. KJ 95T-03, the 100 Talents Program of CAS 
under Contract Nos. U-11, U-24, U-25, and the Knowledge Innovation Project of 
CAS under Contract Nos. U-602, U-34 (IHEP); and by the National Natural Science 
Foundation of China under Contract No. 10175060 (USTC), and No. 10225522 (Tsinghua University).

\end{document}